\documentclass[10pt,a4paper]{article}
\usepackage[utf8]{inputenc}
\usepackage[english]{babel}

\usepackage{amsmath}
\usepackage{amsfonts}
\usepackage{amssymb}

\usepackage[colorlinks,citecolor=blue,urlcolor=blue,linkcolor=blue]{hyperref}

\usepackage[left=2cm,right=2cm,top=2cm,bottom=2cm]{geometry}

\usepackage{authblk}

\newcommand{\pd}{\partial}

\title{Gravity models with nonlinear symmetry realization}

\author[1,2]{S.O. Alexeyev\thanks{\href{mailto:alexeyev@sai.msu.ru}{alexeyev@sai.msu.ru}}}
\author[1,3,4]{D.P. Krichevskiy\thanks{\href{mailto:daniil.krichevskiy@mail.ru}{daniil.krichevskiy@mail.ru}}}
\author[5,6]{B.N. Latosh\thanks{\href{mailto:latosh@theor.jinr.ru}{latosh@theor.jinr.ru}}}

\affil[1]{Sternberg Astronomical Institute, Lomonosov Moscow State University, 119991, Moscow, Russia}
\affil[2]{Department of Quantum Theory and High Energy Physics, Faculty of Physics, Lomonosov Moscow State University, 119991, Moscow, Russia}
\affil[3]{Physics Department, Bauman Moscow State Technical University, Moscow, 105005, Russia}
\affil[4]{Faculty of Science, University of Bern, CH-3012, Bern, Switzerland}
\affil[5]{Bogoliubov Laboratory of Theoretical Physics, JINR, 141980, Dubna, Russia}
\affil[6]{Dubna State University, 141982, Dubna, Russia}

\date{\today}

\begin{document}

\maketitle

\begin{abstract}
  Three models with nonlinear realizations of conformal symmetry are discussed. The simplest model can only describe a universe expanding with a deceleration and does not include inflation. The other models are equivalent up to a variables reparametrization. All these models contain ghost degrees of freedom which may be excluded with an additional symmetry of the target space.
\end{abstract}

\section{Introduction}\label{section_Introduction}

Conformal symmetry plays a particular role in gravity. Its role is defined by the well-known Ogievetsky theorem \cite{Ogievetsky:1973ik}. The theorem states that any generator of the infinitely-dimensional coordinate transformation group is presented as a series of commutators of generators from the conformal group $C(1,3)$ and the affine group $A(4)$. This shows that the conformal group is strongly related with coordinate transformations and shall have a certain influence of gravity theory.

Particular implementations of conformal symmetry for gravity models were studied in multiple papers, so we only mention a few key results. In paper \cite{Borisov:1974bn} in was shown that any metric gravity theory can be viewed as a theory with a combined nonlinear realization of conformal and affine symmetry (see also \cite{Salam:1969rq,Salam:1969bwb,Isham:1970gz,Isham:1971dv}). Explicit nonlinear realization of the conformal symmetry within AdS/CFT approach was constructed in \cite{Bellucci:2002ji}. It also should be mentioned that there are models with a linear realization of the conformal symmetry \cite{Mannheim:2011ds,tHooft:2011aa}, but their applicability is debatable \cite{Riegert:1984hf,Barabash:1999bj,Phillips:2018wao,Caprini:2018oqe}. 

In this paper we extend the discussion started in \cite{Alexeyev:2020lag} of models with nonlinear realizations of conformal symmetry proposed in \cite{Arbuzov:2019rcl}. In the last paper three models were derived. The first one is defined by the Lagrangian:
\begin{align}\label{the_first_model}
  \begin{split}
    \mathcal{L}_I =& \cfrac12 \left[1+\cfrac{\sigma^2}{\varepsilon^2} \,\left\{ f_2\left(\cfrac{\psi}{\varepsilon}\right)\right\}^2\right] ~\eta^{\mu\nu} \,\pd_\mu \psi\,\pd_\nu\psi + \cfrac12 \, \left[ f_1\left(\cfrac{\psi}{\varepsilon}\right)\right]^2 ~ \eta^{\mu\nu} \, \eta_{(\alpha)(\beta)} \, \pd_\mu \sigma^{(\alpha)} \, \pd_\nu \sigma^{(\beta)} \\
    &- \cfrac{1}{\varepsilon} \, f_1\left(\cfrac{\psi}{\varepsilon}\right)\, f_2 \left(\cfrac{\psi}{\varepsilon}\right) ~\eta^{\mu\nu} \, \pd_\mu \psi ~\sigma^{(\alpha)}\pd_\nu \sigma^{(\beta)}\, \eta_{(\alpha)(\beta)}.
  \end{split}
\end{align}
Here $\psi$ and $\sigma^{(\alpha)}$ ($(\alpha)=0,\cdots,4$) are scalar fields associated with target space coordinates on which the conformal symmetry acts nonlinearly. Field indices $\sigma^{(\alpha)}$ taken in brackets should not be confused with Lorentz indices. All fields have the canonical mass dimension and $\varepsilon$ is a mass parameter corresponding to the conformal symmetry breaking scale. Finally, functions $f_i(x)$ are
\begin{align}
  \begin{split}
    f_1(x) &= \cfrac{e^x-1}{x} \,,\\
    f_2(x) &= \cfrac{e^x -x-1}{x^2}\,.
  \end{split}
\end{align}
Therefore the model \eqref{the_first_model} describes five scalar fields propagating in a flat space-time. Fields $\sigma^{(\alpha)}$ have the Minkowski space as the target space so their indices are contracted with the Minkowski target metric $\eta_{(\alpha)(\beta)}$.

The second model is
\begin{align}\label{the_second_model}
  \mathcal{L} =\eta^{\mu\nu} \nabla_\mu h^{(\alpha)(\beta)} \, \nabla_\nu h^{(\rho)(\sigma)} \, \eta_{(\alpha)(\rho)} \, \eta_{(\beta)(\sigma)} ,
\end{align}
where the covariant derivatives are
\begin{align}\label{the_second_model_covariant_derivatives}
  \cfrac{i}{2}\, \nabla_\mu h^{(\alpha)(\beta)} =&\cfrac{i}{2}\, \pd_\mu h^{(\alpha)(\beta)} + \sum\limits_{n=1}^\infty \cfrac{i}{(2n+1)!} \,  (\operatorname{ad}_h^{2n-1} h \pd_\mu h)^{(\alpha)(\beta)}\\
  =& \cfrac{i}{2}\left[  \pd_\mu h^{(\alpha)(\beta)} - \eta_{(\nu)(\sigma)}\eta_{(\mu)(\lambda)}   \left(\cfrac13 \, h^{(\alpha)(\nu)}\, h^{(\beta)(\lambda)} \pd_\mu h^{(\sigma)(\mu)}  -\cfrac13\, h^{(\alpha)(\nu)}\,h^{(\sigma)(\mu)} \pd_\mu h^{(\lambda)(\beta)}\right) + \mathcal{O}(h^5) \right] \,. \nonumber
\end{align}
Here all repeated bracket indices are contracted with the Minkowski metric of the target space and $\operatorname{ad}$ operator is defined as:
\begin{align}
  \left(\operatorname{ad}_h \pd_\mu h\right)^{(\mu)(\nu)}= [h,\pd_\mu h]^{(\mu)(\nu)} = h^{(\mu)(\rho)}\pd_\mu h^{(\sigma)(\nu)} \eta_{(\rho)(\sigma)} - \pd_\mu h^{(\mu)(\rho)} h^{(\sigma)(\nu)} \eta_{(\rho)(\sigma)}.
\end{align}

The third model, in some sense, is similar to the second one:
\begin{align}\label{the_third_model}
  \mathcal{L} =\eta^{\mu\nu} \nabla_\mu h^{(\alpha)(\beta)} \, \nabla_\nu h^{(\rho)(\sigma)} \, \eta_{(\alpha)(\rho)} \, \eta_{(\beta)(\sigma)} + \eta^{\mu\nu} \pd_\mu \phi \pd_\nu \phi.
\end{align}
The same definition of covariant derivatives is used, $\phi$ is a sterile scalar (i.e. a scalar that does not interact with any other degrees of freedom). Further we demonstrate that the second and third models are actually equivalent due to a certain property of the nonlinear symmetry realization used.

The reasons to study these models are the following. Firstly, a nonlinear realization of a symmetry is (almost always) related with a spontaneous symmetry breaking.
In the context of the conformal symmetry this means the following. A physical system admits the conformal symmetry in the high energy phase. In the low energy phase the system develops a conformal symmetry breaking ground state and perturbations propagating about that background are subjected to a nonlinear realization of the symmetry.
Due to the Goldstone theorem after the spontaneous symmetry breaking massless modes appear and some of them could be associated with massless gravitons \cite{Borisov:1974bn}. The key idea is that additional massless modes (for instance, the scalar modes of \eqref{the_first_model}) may be associated with the inflaton. Hence it may be conjectured that the inflation (and the inflaton driving it) is a manifestation of the spontaneously broken conformal symmetry.
Secondly, models \eqref{the_second_model} and \eqref{the_third_model} describe degrees of freedom $h_{\mu\nu}$ that may play a role of tetrads similar to \cite{Borisov:1974bn}. Finally $h_{\mu\nu}$ could contain additional scalar and vector degrees of freedom. Therefore it is interesting to study the content of models \eqref{the_second_model} and \eqref{the_third_model}.

The paper is organized as follows. In Section \ref{section_Inflation} we discuss cosmological regimes described by the model \eqref{the_first_model} and show that scalar degrees of freedom are decoupled. Therefore they do not influence the cosmological behavior in a meaningful way. We discuss a possible relation between this phenomenon and the conformal symmetry together with implications for realistic cosmological scenarios. In section \ref{section_DoF} we discuss the field content of models \eqref{the_second_model} and \eqref{the_third_model}. It is shown that these models actually equivalent up to a coordinate redefinition on the target space. At the same time these models contain vector ghost degrees of freedom. We argue that these degrees of freedom can be excluded via an introduction of an additional symmetry to the target space. However, this symmetry should be agreed with the nonlinear realization of the conformal symmetry which may influence the used nonlinear realization in a meaningful way. Section \ref{section_conclusion} contains our conclusions which extend previous results \cite{Alexeyev:2020lag}. Despite the reasonable motivation proposed in \cite{Arbuzov:2019rcl} they could hardly be realistic without an additional development. The key reason is the presence of ghost degrees of freedom. If ghosts are excluded these models may lose the desirable features justifying their applicability.

\section{Cosmological behavior}\label{section_Inflation}

To study the cosmological behavior of the model \eqref{the_first_model} one has to introduce gravitational degrees of freedom because their number in \eqref{the_first_model} is not enough to describe gravitons. Therefore the standard gravitational degrees of freedom (originating from general relativity (GR)) are required. Such procedure does not contradict the existence of a non-linearly realized conformal symmetry because the used nonlinear realization also generates a linear action of the conformal group on the Lorentz group \cite{Arbuzov:2019rcl}. Therefore to study the cosmological expansion we use the following model:
\begin{align}\label{the_first_model_with_gravity}
  \begin{split}
    \mathcal{S} =& \int d^4 x \sqrt{-g} \Bigg\{-\cfrac{2}{\kappa^2} \, R +\cfrac12 \left[1+\cfrac{\sigma^2}{\varepsilon^2} \,\left\{ f_2\left(\cfrac{\psi}{\varepsilon}\right)\right\}^2\right] ~g^{\mu\nu} \,\pd_\mu \psi\,\pd_\nu\psi \\
    & + \cfrac12 \, \left[ f_1\left(\cfrac{\psi}{\varepsilon}\right)\right]^2 ~ g^{\mu\nu} \, \eta_{(\alpha)(\beta)} \, \pd_\mu \sigma^{(\alpha)} \, \pd_\nu \sigma^{(\beta)} - \cfrac{1}{\varepsilon} \, f_1\left(\cfrac{\psi}{\varepsilon}\right)\, f_2 \left(\cfrac{\psi}{\varepsilon}\right) ~g^{\mu\nu} \, \pd_\mu \psi ~\sigma^{(\alpha)}\pd_\nu \sigma^{(\beta)}\, \eta_{(\alpha)(\beta)} \Bigg\} \,,
  \end{split}
\end{align}
where $\kappa$ is related with the Newton constant $G_N$ as $\kappa^2 = 32 \pi G_N$. Fields $\psi$ and $\sigma^{(\alpha)}$ transform under the nonlinear conformal group action with the target space remaining flat.

Let us consider the cosmological behavior of \eqref{the_first_model_with_gravity} with the open Friedmann space-time:
\begin{align}
  ds^2 = g_{\mu\nu} \, dx^\mu \, dx^\nu = dt^2 - a^2(t) \left( dx^2 + dy^2 + dz^2 \right) .
\end{align}
Here $a(t)$ is the scale factor. The corresponding Einstein equations read
\begin{align}
  \begin{split}
    G_{\mu\nu} = \cfrac{\kappa^2}{4} \, C_{\mu\nu}{}^{\alpha\beta} &\left[\cfrac12\,\left[1+\cfrac{\sigma^2}{\varepsilon^2} \,\left\{ f_2\left(\cfrac{\psi}{\varepsilon}\right)\right\}^2\right]  \,\pd_\alpha \psi\,\pd_\beta\psi+ \cfrac12 \, \left[ f_1\left(\cfrac{\psi}{\varepsilon}\right)\right]^2 ~ \eta_{(\mu)(\nu)} \, \pd_\alpha \sigma^{(\mu)} \, \pd_\beta \sigma^{(\nu)} \right. \\
      &\hspace{15pt}\left.- \cfrac{1}{\varepsilon} \, f_1\left(\cfrac{\psi}{\varepsilon}\right)\, f_2 \left(\cfrac{\psi}{\varepsilon}\right) ~ \pd_\alpha \psi ~\sigma^{(\mu)}\pd_\beta \sigma^{(\nu)}\, \eta_{(\mu)(\nu)}  \right]
  \end{split}
\end{align}
where
\begin{align}
  C_{\mu\nu}{}^{\alpha\beta} \overset{\text{def}}{=} \delta_\mu^\alpha \delta_\nu^\beta+\delta_\mu^\beta\delta_\nu^\alpha- g_{\mu\nu}g^{\alpha\beta} \,.
\end{align}
These equations have two non-vanishing components which can be reduced to:
\begin{align}\label{Cosmological_equations}
  \begin{split}
    &-3\cfrac{\ddot{a} }{a} =\cfrac{1}{4} \, \kappa^{2} \, \left( \cfrac12\,\left[1+\cfrac{\sigma^2}{\varepsilon^2} \,\left\{ f_2\left(\cfrac{\psi}{\varepsilon}\right)\right\}^2\right] \dot{\psi }^{2} + \left[ f_1\left(\cfrac{\psi}{\varepsilon}\right)\right]^2 \dot\sigma^{(\alpha)} \dot\sigma_{(\alpha)} -\frac{2}{\varepsilon} f_{1}\left( \frac{\psi }{\varepsilon} \right)  f_{2}\left( \frac{\psi }{\varepsilon} \right)\dot\psi\,  \dot{\sigma }_{(\alpha )} \sigma^{(\alpha )} \right)  \, 
  ,\\
    &2\dot{a}^{2} +a\ddot{a}=0 \,.
  \end{split}
\end{align}
The second equation from \eqref{Cosmological_equations} does not contain scalar fields and describes the behavior of the scalar factor by itself. It can be solved analytically:
\begin{equation}\label{scale_factor} 
    a(t)=c_2 \left( c_1+3\, t \right)^{\frac13}\, ,
\end{equation}
where $c_1$ and $c_2$ are the integration constants defined by the boundary conditions.

Now we analyze the result \eqref{scale_factor}. First of all we note that here the universe has only a decelerated expansion hence the model has no room for an inflationary phase.

Secondly, one can assume that the matter content of the model (i.e. the scalar fields) is described by the standard equation of state (EoS) $p = w \rho$. Here $p$ is a pressure of scalar fields, $\rho$ is an energy density of the matter, and $w$ is the EoS parameter. Solutions with such EoS are well known \cite{Weinberg,Gorbunov:2011zz}, so one can easily restore $w$ from the form of asymptotic of \eqref{scale_factor}. At large values of time the scale factor is proportional to $t^{\frac13}$ corresponding to EoS parameter $w=1$. That is why despite the fact that scalar fields in this model admit vanishing masses their behavior do not correspond to a relativistic matter one with EoS parameter $w=1/3$. The reason is that the discussed scalar fields have a non-trivial interaction sector which influence their EoS.

Finally, the fact that the model admits a decelerating solution deserves a special attention as it has ghost degrees of freedom. These ghost degrees of freedom appear due to the metric on scalar field target space $\eta_{(\mu)(\nu)}$:
\begin{align}
  g^{\mu\nu} \, \eta_{(\alpha)(\beta)} \, \pd_\mu \sigma^{(\alpha)} \, \pd_\nu \sigma^{(\beta)} = \pd^\mu \sigma^{(0)} \, \pd_\mu \sigma^{(0)}  - \sum\limits_{i=1}^3 \pd^\mu \sigma^{(i)} \, \pd_\mu \sigma^{(i)} \,.
\end{align}
Note that these ghost degrees of freedom do not manifest themselves at the level of cosmological solutions.

Despite the fact the an analytical solution \eqref{scale_factor} can be obtained, there are no reasons to believe that it is stable. Equations \eqref{Cosmological_equations} take a simple form due to a cancellation of the stress-energy tensor of scalar fields. This cancellation, in turn, is possible because both metric and scalar fields depend only on the time variable. As soon as one considers metric and scalar field perturbations propagating around the background a similar cancellation becomes impossible. 

The existence of ghost degrees of freedom makes possible implementations of \eqref{the_first_model_with_gravity} unlikely. Let us discuss possible opportunities to exclude ghosts. The first simple way is to use the inverse Higgs mechanism \cite{Ivanov:1975zq}. Unfortunately this procedure is not applicable here because the discussed nonlinear symmetry realization does no satisfy the necessary criteria. Another opportunity to exclude ghosts is to introduce an additional symmetry at the scalar field target space which would make $\sigma^{(1)} = \sigma^{(2)} = \sigma^{(3)} = 0$. Consequently, only a single massless sterile scalar field (i.e. it has neither self-interaction nor potential) $\psi$ remains and such a field can hardly be applied in realistic scenarios. The best opportunity wpuld be to find a mechanism excluding ghost degrees of freedom from the model's physical spectrum, but allowing them to propagate only in loops. As it was shown in \cite{Arbuzov:2020swg} for such a case the scalar field $\psi$ develops non-trivial interaction at the loop level. However, such a mechanism has yet to be found.

Summarizing, we conclude that the model \eqref{the_first_model} can hardly be applied for realistic scenarios. The main issue is the existence of ghost degrees of freedom that expectidly lead to an instability. Even if one finds a way to negate the ghost problem the known cosmological solutions \eqref{scale_factor} show that the scalar fields act as matter with EoS parameter $w=1$ and cannot drive an inflation. Therefore one shall consider the model \eqref{the_first_model} unrealistic.

\section{Field content}\label{section_DoF}

Now we switch to a discussion of the field content of second \eqref{the_second_model} and third \eqref{the_third_model} models. Firstly, they have at least ten degrees of freedom from the symmetric matrix $h^{(\alpha)(\beta)}$, therefore they can describe spin-$2$ massless degrees of freedom associated with gravitons. Secondly, $h^{(\alpha)(\beta)}$ may also contain two scalar degrees of freedom (associated with its determinant and trace) which could serve as inflatons. Finally, the third model \eqref{the_third_model} contains a sterile scalar appearing as a consequence of the properties of operator $D$ of the conformal group \cite{Arbuzov:2019rcl}. These issues are clarified further.

First and foremost we shall address the issue related with the sterile scalar of model \eqref{the_third_model}. In the original article \cite{Arbuzov:2019rcl} it was missed that the sterile scalar is related with the trace of $h^{(\alpha)(\beta)}$. The reason behind this is due to the relation between the conformal group generators.
Namely, as degrees of freedom $h^{(\alpha)(\beta)}$ are associated with the coset coordinates along $R_{(\alpha)(\beta)}$ direction the sterile scalar $\phi$ is associated with the coset coordinates along $D$ direction also. However operators $R_{(\alpha)(\beta)}$ are dependent and coupled as \cite{Ogievetsky:1973ik,Borisov:1974bn,Arbuzov:2019rcl}:
\begin{align}\label{eq:R_2D}
  \eta^{(\alpha)(\beta)}\, R_{(\alpha)(\beta)} = 2 D \, .
\end{align}
Therefore the corresponding coset coordinates are also dependent and the trace $\eta_{(\alpha)(\beta)} h^{(\alpha)(\beta)}$ should be associated with $\phi$. Hence the trace component of $h^{(\alpha)(\beta)}$ acts as a sterile massless scalar and therefore cannot drive the inflation. Moreover, one can treat $h^{(\alpha)(\beta)}$ as a traceless matrix with $9$ independent components reducing the number of valuable degrees of freedom in the model. On the other hand it guarantees the traceless of $h^{(\alpha)(\beta)}$ similar to GR degrees of freedom. 

Finally, note that this result can be obtained independently via the direct verification. The definition $h^{(\alpha)(\beta)}= \cfrac14 \,\eta^{(\alpha)(\beta)} \, h $ with $h=\eta_{(\alpha)(\beta)} h^{(\alpha)(\beta)}$ causes  the covariant derivative \eqref{the_second_model_covariant_derivatives} to be completely reduced to the regular ones hence the Lagrangian \eqref{the_second_model} describes only a single massless sterile scalar.

Now we switch to $\overline{h}^{(\alpha)(\beta)}$ as the traceless part of $h^{(\alpha)(\beta)}$ and start to study its Lagrangian. The corresponding covariant derivative \eqref{the_second_model_covariant_derivatives} is:
\begin{align}
  \begin{split}
    &\cfrac{i}{2} \,\nabla_\mu \overline{h}^{(\alpha)(\beta)}\\
    &=\cfrac{i}{2}\left[  \pd_\mu \overline{h}^{(\alpha)(\beta)} - \eta_{(\nu)(\sigma)}\eta_{(\mu)(\lambda)}   \left(\cfrac13 \, \overline{h}^{(\alpha)(\nu)}\, \overline{h}^{(\beta)(\lambda)} \pd_\mu \overline{h}^{(\sigma)(\mu)}  -\cfrac13\, \overline{h}^{(\alpha)(\nu)}\,\overline{h}^{(\sigma)(\mu)} \pd_\mu \overline{h}^{(\lambda)(\beta)}\right) + \mathcal{O}(h^5) \right]\,.
  \end{split}
\end{align}
This derivative matches the expression \eqref{the_second_model_covariant_derivatives} because the trace component is contained only in $\pd_\mu h^{(\alpha)(\beta)}$. Therefore the expression \eqref{the_second_model} determines the traceless part of $h^{(\alpha)(\beta)}$ without changing its form.

Further as the target space metric is not positively defined the model may contain ghosts. Let us demonstrate this explicitly. The Lagrangian density $\mathcal{L}$ of \eqref{the_second_model} given up to $\mathcal{O}(h^3)$ reads:
\begin{align}
  \begin{split}
    \mathcal{L} =&\cfrac{1}{2} \, \eta_{(\alpha)(\rho)}\eta_{(\beta)(\sigma)} \, \pd_\mu \overline{h}^{(\alpha)(\beta)} \, \pd^\mu \overline{h}^{(\rho)(\sigma)}\\
    =& \cfrac12 \, \pd_\mu \overline{h}^{(0)(0)} \pd^\mu \overline{h}^{(0)(0)} + \cfrac12 \, \sum\limits_{a,b=1}^3\, \pd_\mu \overline{h}^{(a)(b)} \,\pd^\mu \overline{h}^{(a)(b)} -  \sum\limits_{s=1}^3 \, \pd_\mu \overline{h}^{(0)(s)} \, \pd^\mu \overline{h}^{(0)(s)}\,.
  \end{split}
\end{align}
It is clearly seen that $\overline{h}^{(0)(s)}$ with $s=1,2,3$ have the wrong sign kinetic term and describe ghost degrees of freedom. Because of the off-diagonal elements $h^{(0)(i)}$ the model cannot be considered realistic until these degrees of freedom are excluded. These ghosts cannot be excluded via the inverse Higgs mechanism as the corresponding generators do not satisfy the required conditions \cite{Ivanov:1975zq}.

One can introduce an additional symmetry in the target space. The choice of the target space coordinates as $\zeta^{(\alpha)}$ causes the following form of the target space metric:
\begin{align}
  g = \eta_{(\alpha)(\beta)} \, d\zeta^{(\alpha)}\otimes d\zeta^{(\beta)}\,.
\end{align}
If one demands a symmetry with respect to $\zeta^{(0)}$ inversion ($\zeta^{(0)} \to -\zeta^{(0)}$) then the field $h^{(\alpha)(\beta)}$ must have the vanishing component $h^{(0)(i)}$ to be consistent with the considering symmetry. This symmetry condition should be considered independently from the particular choice of the nonlinear conformal symmetry realization. However, it is unclear what physical reason can justify the introduction of such a symmetry. Therefore for the time being we consider it as an ad hoc prescription.

With the discussed prescription the traceless part of $h^{(\alpha)(\beta)}$ includes six independent components which can fit one spin-$2$ degree of freedom and one spin-$0$ degree of freedom. To obtain the vanishing trace we set
\begin{align}
  \overline{h}^{(0)(0)} = \sum\limits_{a=1}^3 \overline{h}^{(a)(a)} .
\end{align}
In such a presentation the only non-vanishing components are $h^{(a)(b)}$ where $a,b=1,2,3$. Within such a setup the original Lagrangian \eqref{the_second_model} up to $\mathcal{O}(h^4)$ order looks as:
\begin{align}
  \begin{split}
    \mathcal{L} =&\pd_\mu h^{(1)(1)} \,\pd^\mu h^{(1)(1)}+\pd_\mu h^{(2)(2)} \,\pd^\mu h^{(2)(2)}+\pd_\mu h^{(3)(3)} \,\pd^\mu h^{(3)(3)}\\
    &+\pd_\mu h^{(1)(2)} \,\pd^\mu h^{(1)(2)}+\pd_\mu h^{(2)(3)} \,\pd^\mu h^{(2)(3)}+\pd_\mu h^{(3)(1)} \,\pd^\mu h^{(3)(1)} \\
    &\pd_\mu h^{(1)(1)} \,\pd^\mu h^{(2)(2)}+\pd_\mu h^{(2)(2)} \,\pd^\mu h^{(3)(3)}+\pd_\mu h^{(3)(3)} \,\pd^\mu h^{(1)(1)}  +\mathcal{O}(h^4) \,.
  \end{split}
\end{align}
Generically this Lagrangian is non-diagonal. It could be made diagonal in the following representation: 
\begin{align}
  \begin{split}
    \zeta_1 =& \cfrac{1}{\sqrt{6}} \,\left[h^{(1)(1)}+h^{(2)(2)}+h^{(3)(3)}\right] \,, \\
    \zeta_2 =& h^{(1)(1)} -  h^{(3)(3)} \, ,\\
    \zeta_3 =& h^{(1)(1)} -  h^{(2)(2)} \, ,
  \end{split}
  &
  \begin{split}
    \zeta_4 =& h^{(1)(2)}\,, \\
    \zeta_5 =& h^{(2)(3)}\,, \\
    \zeta_6 =& h^{(3)(1)}\,.
  \end{split}
\end{align}
The quadratic part of the diagonal Lagrangian reads
\begin{align}
  \mathcal{L} =\sum\limits_{i=1}^6 \pd_\mu \zeta_i \,\pd^\mu \zeta_i
\end{align}
so the corresponding field equations are reduced to the Klein-Gordon equation
\begin{align}
  \square \, \zeta_i =0\,.
\end{align}
As a result the model describes massless degrees of freedom. However one lacks a condition which can fix their chirality.

Mass and chirality of a particle are fixed by eigenvalues of the Poincare group Casimir operators \cite{Wigner:1939cj,Bilal:2001nv}. The D'Alamber operator $\square$ defined to the mass operator which is one of the two Poincare group Casimir operators. The second Casimir operator is the Pauli-Lubanski vector
\begin{align}
  W^\mu = \cfrac12 \, \varepsilon^{\mu\nu\alpha\beta} \, P_\nu\, M_{\alpha\beta}\,
\end{align}
where $M_{\alpha\beta}$ are Lorentz generators building  Lorentz group action on a given degree of freedom. The degrees of freedom $h^{(\alpha)(\beta)}$ should be mixed to diagonalize the Lagrangian therefore they can be only scalars of the trivial Lorenz group action. Hence the model has no room for spin-$2$ degrees of freedom.

This section could be summarized as follows. Both discussed models \eqref{the_second_model} and \eqref{the_third_model} really are equal because of the relations between the conformal group generators with the influence on the coset coordinates and the corresponding nonlinear symmetry realization. Secondly, the model \eqref{the_second_model} contains ghost degrees of freedom that can be excluded from the model only with an ad hoc prescription of auxiliary symmetry of the model target space. Such a prescription makes the model to be not natural. Nonetheless, the model with the additional symmetry is healthy and describes massless degrees of freedom appearing to be scalar ones. All these factors make the considered model not realistic.

\section{Discussion and conclusions}\label{section_conclusion}

Three models with particular nonlinear conformal symmetry realizations \cite{Arbuzov:2019rcl} were studied. We extend the consideration started in \cite{Alexeyev:2020lag} and demonstrate that the discussed models could become realistic only after significant modifications.

The first model \eqref{the_first_model} seems not to be realistic. Firstly, the original model contains five degrees of freedom propagating in a flat space-time. In \cite{Arbuzov:2019rcl} it was argued that an extension to a curved space-time case may realize a viable inflationary scenario. Following \cite{Alexeyev:2020lag} we show that in this model a universe expands with a deceleration (EoS parameter $w=1$) and, therefore, there is no inflation in it. Secondly, the model contains ghost degrees of freedom. Although they do not appear at the cosmological solution it is reasonable to expect that they make the solution unstable. Therefore the model is not applicable unless ghosts are excluded from the perturbation spectrum.

Secondly, \eqref{the_second_model} and \eqref{the_third_model} were analyzed and found out to be equivalent. This feature was missed both in \cite{Arbuzov:2019rcl} and \cite{Alexeyev:2020lag}. The models are equivalent up to a parametrization of dynamical variables. As it is pointed in the previous section the trace component of coset coordinates $h^{(\alpha)(\beta)}$ present in model \eqref{the_second_model} is the coset coordinate $\phi$ present in  model \eqref{the_third_model}. The coset coordinate $\phi$ is associated with operator $D$ which is related with the trace of operators $R_{(\alpha)(\beta)}$ which, in turn, are associated with coordinated $h_{(\alpha)(\beta)}$ (see \eqref{eq:R_2D}). Therefore models \eqref{the_second_model} and \eqref{the_third_model} merely provide different parametrizations of the same model.

Finally, we analyzed the field content of model \eqref{the_second_model}. The trace component of $h_{(\alpha)(\beta)}$ acts as a sterile massless scalar which excludes its practical application. The traceless part $\overline{h}_{(\alpha)(\beta)}$ contains nine degrees of freedom. They are three ghosts $\overline{h}^{(0)(s)}$ with $s=1,2,3$ which cannot be excluded with the inverse Higgs mechanism\cite{Ivanov:1975zq}. An additional symmetry of the target space may exclude the ghosts. Therefore an opportunity to introduce such a symmetry should be studied.

\section*{Acknowledgment}
The work (B.L.) was supported by the Foundation for the Advancement of Theoretical Physics and Mathematics “BASIS”. The work (S.A.) was also supported by the Interdisciplinary Scientific and Educational School of Moscow University ``Fundamental and Applied Space Research''.

\section*{Appendix A}

Here it is necessary to demonstrate some features of an interaction between $h^{(\mu)(\nu)}$ and matter degrees of freedom within model \eqref{the_second_model}. The issue provides an additional reason to believe that the model can hardly be considered realistic.

For the sake of simplicity we consider a single massless vector field (which can be associated with the electromagnetic field). Accordingly to \cite{Arbuzov:2019rcl} such a vector field is described by the standard Lagrangian
\begin{align}
  \mathcal{L}_\text{vec} = -\cfrac14\,F_{\mu\nu} F^{\mu\nu} \, .
\end{align}
However, a definition of the field tensor contains covariant derivatives:
\begin{align}
  F_{\mu\nu} = \nabla_\mu A_\nu - \nabla_\nu A_\mu \, .
\end{align}
The covariant derivative, in turn, reads
\begin{align}
  \begin{split}
    \nabla_{\mu } A^{\nu }=\partial_{\mu } A^{\nu }+\sum^{\infty }_{n=0} \frac{i}{(2n+2)!} (ad^{2n}_{h}h\partial_{\mu } h)^{(\alpha )(\beta )}(M_{(\alpha )(\beta )})^{\nu }_{\sigma }A^{\sigma } \\
    =  \partial_{\mu } A^{\nu }+\frac{i}{2} \eta_{(\gamma )(\sigma )} h^{(\alpha )(\gamma )} \partial_{\mu } h^{(\beta )(\sigma )} (M_{(\alpha )(\beta )})^{\nu }_{\sigma }A^{\sigma }+\mathcal{O}(h^{2} ) \, .
  \end{split}
\end{align}
Here $\left(M_{(\alpha)(\beta)}\right)_{\mu\nu} = i  (\eta_{(\alpha)\mu} \eta_{(\beta)\nu} - \eta_{(\alpha)\nu} \eta_{(\beta)\mu})$ is the vector representation of Lorentz group generators.

The corresponding Lagrangian has a few peculiar features. First and foremost, at the leading order the model describes interaction between two modes $h^{(\mu)(\nu)}$ and two vectors $A^\mu$. Within GR, in contrast, the leading order interaction between gravity and a massless vector field is a cubic interaction describing an interaction between two vectors and a graviton. Secondly, in contrast with GR the interaction can contain only odd powers of $h^{(\mu)(\nu)}$. 

Implications of these features lead to significant differences between GR and \eqref{the_second_model}. The most interesting one is related with $2\to 2$ scattering. Within GR there are tree-level amplitudes describing gravitational scattering of massless vectors. Within model \eqref{the_second_model} such a scattering appears only at the one-loop level.

It is necessary to point out that to pinpoint the exact difference in such scattering processes a more detailed analysis is required. However, even at the current level the model \eqref{the_second_model} appears to be inconsistent with GR.

\bibliographystyle{unsrt}
\bibliography{Main.bib}

\begin{thebibliography}{10}

\bibitem{Ogievetsky:1973ik}
V.~I. Ogievetsky.
\newblock {Infinite-dimensional algebra of general covariance group as the
  closure of finite-dimensional algebras of conformal and linear groups}.
\newblock {\em Lett. Nuovo Cim.}, 8:988--990, 1973.

\bibitem{Borisov:1974bn}
A.~B. Borisov and V.~I. Ogievetsky.
\newblock {Theory of Dynamical Affine and Conformal Symmetries as Gravity
  Theory}.
\newblock {\em Theor. Math. Phys.}, 21:1179, 1975.

\bibitem{Salam:1969rq}
Abdus Salam and J.~A. Strathdee.
\newblock {Nonlinear realizations. 1: The Role of Goldstone bosons}.
\newblock {\em Phys. Rev.}, 184:1750--1759, 1969.

\bibitem{Salam:1969bwb}
Abdus Salam and J.~A. Strathdee.
\newblock {Nonlinear realizations. 2. Conformal symmetry}.
\newblock {\em Phys. Rev.}, 184:1760--1768, 1969.

\bibitem{Isham:1970gz}
C.~J. Isham, Abdus Salam, and J.~A. Strathdee.
\newblock {Spontaneous breakdown of conformal symmetry}.
\newblock {\em Phys. Lett.}, 31B:300--302, 1970.

\bibitem{Isham:1971dv}
C.~J. Isham, Abdus Salam, and J.~A. Strathdee.
\newblock {Nonlinear realizations of space-time symmetries. Scalar and tensor
  gravity}.
\newblock {\em Annals Phys.}, 62:98--119, 1971.

\bibitem{Bellucci:2002ji}
S.~Bellucci, E.~Ivanov, and S.~Krivonos.
\newblock {AdS / CFT equivalence transformation}.
\newblock {\em Phys. Rev. D}, 66:086001, 2002.
\newblock [Erratum: Phys.Rev.D 67, 049901 (2003)].

\bibitem{Mannheim:2011ds}
Philip~D. Mannheim.
\newblock {Making the Case for Conformal Gravity}.
\newblock {\em Found. Phys.}, 42:388--420, 2012.

\bibitem{tHooft:2011aa}
Gerard 't~Hooft.
\newblock {A class of elementary particle models without any adjustable real
  parameters}.
\newblock {\em Found. Phys.}, 41:1829--1856, 2011.

\bibitem{Riegert:1984hf}
R.~J. Riegert.
\newblock {The particle content of linearized conformal gravity}.
\newblock {\em Phys. Lett. A}, 105:110--112, 1984.

\bibitem{Barabash:1999bj}
O.~V. Barabash and Yu.~V. Shtanov.
\newblock {Newtonian limit of conformal gravity}.
\newblock {\em Phys. Rev. D}, 60:064008, 1999.

\bibitem{Phillips:2018wao}
Peter~R. Phillips.
\newblock {Schwarzschild and linear potentials in Mannheim\textquoteright{}s
  model of conformal gravity}.
\newblock {\em Mon. Not. Roy. Astron. Soc.}, 478(2):2827--2834, 2018.

\bibitem{Caprini:2018oqe}
Chiara Caprini, Patric H\"olscher, and Dominik~J. Schwarz.
\newblock {Astrophysical Gravitational Waves in Conformal Gravity}.
\newblock {\em Phys. Rev. D}, 98(8):084002, 2018.

\bibitem{Alexeyev:2020lag}
S.~Alexeyev and D.~Krichevskiy.
\newblock {Inflationary solutions in the simplest gravity model with conformal
  symmetry}.
\newblock {\em Phys. Part. Nucl. Lett.}, 18(2):128--130, 2021.

\bibitem{Arbuzov:2019rcl}
A.~B. Arbuzov and B.~N. Latosh.
\newblock {Gravity and Nonlinear Symmetry Realization}.
\newblock {\em Universe}, 6(1):12, 2020.

\bibitem{Weinberg}
Steven Weinberg.
\newblock {\em Cosmology}.
\newblock Oxford University Press, 2008.

\bibitem{Gorbunov:2011zz}
Valery~A. Rubakov and Dmitry~S. Gorbunov.
\newblock {\em {Introduction to the Theory of the Early Universe}}.
\newblock World Scientific, Singapore, 2017.

\bibitem{Ivanov:1975zq}
E.~A. Ivanov and V.~I. Ogievetsky.
\newblock {The Inverse Higgs Phenomenon in Nonlinear Realizations}.
\newblock {\em Teor. Mat. Fiz.}, 25:164--177, 1975.

\bibitem{Arbuzov:2020swg}
Andrej Arbuzov and Boris Latosh.
\newblock {On anomalies in effective models with nonlinear symmetry
  realization}.
\newblock {\em Mod. Phys. Lett. A}, 35(35):2050294, 2020.

\bibitem{Wigner:1939cj}
Eugene~P. Wigner.
\newblock {On Unitary Representations of the Inhomogeneous Lorentz Group}.
\newblock {\em Annals Math.}, 40:149--204, 1939.
\newblock [Reprint: Nucl. Phys. Proc. Suppl.6,9(1989)].

\bibitem{Bilal:2001nv}
Adel Bilal.
\newblock Introduction to supersymmetry.
\newblock 2001.

\end{thebibliography}

\end{document}